%% file: amp2e.tex
\documentclass[12pt]{article}

\usepackage{amsmath}
\usepackage{amssymb}
\usepackage{epsfig}
\usepackage{graphicx}
\usepackage{color}
\usepackage{xspace}
\usepackage{url}
\usepackage{layout}

\newcommand{\proofend}{\nopagebreak \hfill
{\framebox{\rule{.5ex}{0ex}\rule{0ex}{.5ex}}} \par}

\newcounter{figurecount}

\def\mb{$\begin{displaystyle}}
\def\me{\end{displaystyle}$\ }

\newcommand{\R}{{\rm \bf R}}

\newcommand{\C}{{\rm \bf C}}

\def\glb{\mbox{\rm glb}}

\DeclareMathOperator{\tr}{tr}

\newcommand{\ket}[1]{\left| #1 \right\rangle}
\newcommand{\bra}[1]{\left\langle #1 \right|}

\newcommand{\Stab}{\mbox{\rm Stab}}
\DeclareMathOperator{\Sym}{Sym}

\begin{document}

\thispagestyle{empty}

\title{Werner State Structure and Entanglement Classification}

\author{David W. Lyons\footnote{Mathematical Sciences, Lebanon Valley
    College, PA USA} \and Abigail M. Skelton\footnote{Mathematical
    Sciences, Lebanon Valley
    College, PA USA}  \and Scott N. Walck\footnote{Physics, Lebanon Valley
    College, PA USA} }

\date{28 September 2011, revised 3 April 2012}


\maketitle

\begin{abstract}
We present applications of the representation theory of Lie groups to
the analysis of structure and local unitary classification of Werner
states, sometimes called the {\em decoherence-free} states, which are states
of $n$ quantum bits left unchanged by local transformations that are the
same on each particle. We introduce a multiqubit generalization of the
singlet state, and a construction that assembles these into Werner
states.
\end{abstract}

\section{Introduction}

Quantum entanglement, a feature of quantum theory named by
Schr\"odinger~\cite{schroedingercoinentanglement} and employed by
Bell~\cite{bell64} in the rejection of local realism, has come to be
seen as a resource for quantum information processing tasks including
measurement-based quantum computation, teleportation, and some forms of
quantum cryptography. Driven by applications to computation and
communication, entanglement of composite systems of $n$ quantum bits, or
{\em qubits}, is of particular interest.

The problem of entanglement is to understand nonlocal properties of
states and to answer operational questions such as when two given states
can be interconverted by local operations on individual subsystems. This
inspires the mathematical problem of classifying orbits of the local
unitary group action on the space of states. 

The goal of this article is to address these questions for the Werner
states, which are defined to be those states invariant under the action
of any particular single qubit unitary operator acting on all $n$
qubits. Werner states have found a multitude of uses in quantum
information science.  Originally introduced in 1989 for two
particles~\cite{werner89} to distinguish between classical correlation
and Bell inequality satisfaction, Werner states have found use in the
description of noisy quantum channels~\cite{lee00}, as examples in
nonadditivity claims~\cite{shor01}, and in the study of deterministic
purification~\cite{short09}.  In what may prove to be a practical
application to computing in noisy environments, Werner states lie in the
decoherence-free subspace for collective
decoherence~\cite{zanardi97,lidar98,bourennane04}. A recent example of how
analysis of state structure can be useful is the work of Migda{\l} et al.~\cite{migdal2011} 
on protecting information against the loss of a qubit using
Werner states.

We apply the representation theory of Lie groups, in particular,
Clebsch-Gordan decomposition of representations of $SU(2)$ on tensor
products and the representation theory of $SO(3)$ on polynomials in
three variables, to obtain structural theorems and local unitary
classification for Werner states.  We summarize recent results for the
special cases of {\em pure} Werner states~\cite{su2blockstates} and {\em
  symmetric} Werner states~\cite{symmwerner} in
Section~\ref{previousresults} below. We present new results for the
general case of mixed Werner states in
Section~\ref{mixedwernergeneral}. We introduce a generalization of the
singlet state and use these states to construct Werner states.

\section{Local Unitary Group Action}
\label{luact}

Let $G=(SU(2))^{n}$ denote the local unitary (LU) group for
$n$-qubit states. An LU operator $g=(g_1,g_2,\ldots,g_n)$ acts on an $n$-qubit
density matrix $\rho$ (that is, a $2^n\times 2^n$ positive semidefinite
matrix with trace 1) by
$$\rho\mapsto g\rho g^\dagger:=(g_1\otimes g_2\otimes\cdots \otimes g_n) \rho
(g_1^\dagger\otimes g_2^\dagger\otimes\cdots \otimes g_n^\dagger).
$$
In this notation, the Werner states are defined to be the set of density
matrices $\rho$ such that $\rho = g^{\otimes n} \rho(g^\dagger)^{\otimes
  n}$ for all $g$ in $SU(2)$. We will write $\Delta$ to denote the
subgroup
$$\Delta = \{(g,g,\ldots,g)\colon g\in SU(2)\}$$
of the LU group $G$.

The set of $n$-qubit density matrices is a convex set inside of the
vector space ${\cal V}^{\otimes n}$, where ${\cal V}$ is the 4-dimensional real
vector space of $2 \times 2$ Hermitian matrices.  A convenient basis for
${\cal V}$ is $\{ \sigma_0, \sigma_1, \sigma_2, \sigma_3 \}$, where $\sigma_0$
is the $2 \times 2$ identity matrix, and $\sigma_1=\sigma_x$,
$\sigma_2=\sigma_y$, and $\sigma_3=\sigma_z$ are the Pauli matrices.
Every element $\rho$ (whether or not $\rho$ is
positive or has trace 1) of ${\cal V}^{\otimes n}$ can be uniquely written in the form $\rho=\sum_I
s_I\sigma_I$, where $I=i_1i_2\ldots i_n$ is a multiindex with
$i_k=0,1,2,3$ for $1\leq k\leq n$, and $\sigma_I$ denotes
$$\sigma_I = \sigma_{i_1}\otimes \sigma_{i_2}\otimes \cdots \otimes
\sigma_{i_n},$$ with real coefficients $s_I$.

Sitting inside ${\cal V}^{\otimes n}$ is the space of pure states, which
are the rank 1 density matrices of the form $\ket{\psi}\bra{\psi}$,
where $\ket{\psi}$ is a vector in the Hilbert space ${\cal
  H}=(\C^2)^{\otimes n}$ of pure $n$-qubit states. We will use the
computational basis vectors $\ket{I}$ for ${\cal H}$, where $I=
i_1i_2\ldots i_n$ is a multiindex with $i_k=0,1$ for $1\leq k\leq
n$. The expansion of a pure state vector $\ket{\psi}$ in the
computational basis has the form $\ket{\psi} = \sum_I c_I \ket{I}$,
where the coefficients $c_I$ are complex. Note that we use the same
multiindex notation $I$ for ``mod 4'' multiindices for tensors of Pauli
matrices in ${\cal V}^{\otimes n}$, and for ``mod 2'' multiindices for
computational basis vectors in ${\cal H}$. The distinction will be clear
from context.

\section{Pure and Symmetric Werner States}
\label{previousresults}

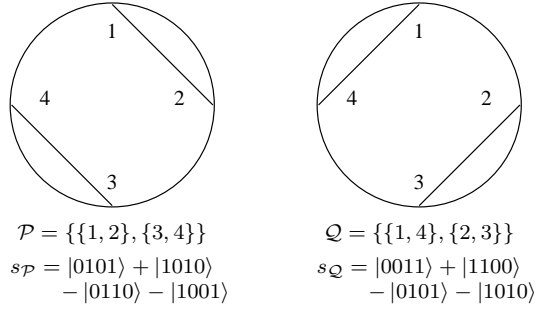
\begin{figure}
  \begin{center}
\input{spillustration4.pstex_t}
\end{center}
\caption{\label{spillustrationfig}The two non-intersecting 4-qubit chord
  diagrams and their associated singlet product states.}
\end{figure}

In~\cite{su2blockstates}, we prove a structure theorem for pure Werner
states based on the following geometric construction. Begin with a
circle with an even number $n=2m$ of marked points, labeled 1 through $n$
in order around the circle, say clockwise. Let ${\cal P}$ be a partition
of $\{1,2,\ldots,n\}$ into two-element subsets. Each two-element subset
$\{a,b\}$ determines a chord connecting $a$ and $b$. We impose the
condition that no two chords coming from ${\cal P}$ may intersect. For
each chord $C$ in ${\cal P}$, let $\ket{s_C}$ be a singlet state
$\frac{1}{\sqrt{2}}(\ket{01}-\ket{10})$ in the two qubits at the ends of
$C$, and define the state $\ket{s_{\cal P}}$ to be the product 
$$\ket{s_{\cal P}}=\bigotimes_{C\in{\cal P}} \ket{s_C}$$
of singlet
states $\ket{s_C}$, over all $C$ in ${\cal P}$. We call states of the
form $\ket{s_{\cal P}}$ ``non-intersecting chord diagram states''. Figure~\ref{spillustrationfig} illustrates
the two possibilities for 4 qubits.

We show that any linear combination of chord diagram states is a Werner
state, and conversely, any pure Werner state can be written uniquely as
a linear combination of non-intersecting chord diagram states. Further,
these linear combinations are unique representatives of their LU
equivalence class, up to a phase factor. Representation theory and
combinatorics enter the story in the proof that the non-intersecting
chord diagram states span the space of pure Werner states. The Werner
states are the trivial summand in the decomposition into irreducible
submodules of the $SU(2)$-space ${\cal H}=(\C^2)^{\otimes n}$. The
dimension of the trivial summand is equal to the Catalan number
$$\frac{1}{m+1}{2m \choose m}$$ when the number of qubits $n=2m$ is
even, and the dimension of this space is zero when $n$ is odd. The
non-intersecting chord diagrams with $n=2m$ nodes are one of the
well-known sets enumerated by the Catalan
numbers~\cite{stanleyenumcombvol2}. Together with an argument that the
non-intersecting chord diagram states are linearly independent, the fact
that these two numbers agree establishes that the chord diagram states
form a linear basis for the space of pure Werner states.

In~\cite{symmwerner}, we consider the case of pure and mixed Werner
states that are invariant under permutations of qubits, also called
{\em symmetric} states. Given nonnegative integers $n_1,n_1,n_3$ with
$n_1+n_2+n_3\leq n$, we identify the monomial $x^{n_1}y^{n_2}z^{n_3}$
in three variables with the matrix $$\rho = \alpha \Sym (\sigma_0^{\otimes
  n_0} \otimes \sigma_1^{\otimes n_1}\otimes \sigma_2^{\otimes
  n_2}\otimes \sigma_3^{\otimes n_3})$$
where $n_0=n-n_1-n_2-n_3$, the symmetrizing operator $\Sym$ sums all the permutations
of the products of $n_k$ copies of $\sigma_k$ for $k=0,1,2,3$, and
$\alpha$ is a normalization factor. This establishes a correspondence
between mixed symmetric states (not necessarily Werner states) and real
polynomials in three variables. Using the representation theory of
$SO(3)$, we show that symmetric Werner states correspond to polynomials
that are linear combinations of $(x^2+y^2+z^2)^m$ for some $m\leq
\lfloor n/2\rfloor$. Further, any two such states are local unitarily
inequivalent. 

Now we turn to the general case of mixed Werner states.

\section{The General Case of Mixed Werner States}
\label{mixedwernergeneral}

We begin with the construction of a family of density matrices $C_n$
that generalize the singlet state.

Given an $n$-qubit binary string $I$, let $C(I)$ denote the pure state
$$C(I) = \alpha \sum_{k=0}^{n-1} \omega^k  \ket{\pi^k I}
$$ where $\omega=e^{\frac{2\pi i}{n}}$ and $\pi$ is the the 
cyclic permutation of $\{1,2,\ldots,n\}$ given by $1\mapsto n$,
$k\mapsto k-1$ for $2\leq k\leq n$, and $\alpha$ is a normalizing factor
so that $|C(I)|=1$, whenever $C(I)\neq 0$ (notice that $C(00)=0$, so is not a
state). For example, 
$$C(001) = \frac{1}{\sqrt{3}}(\ket{001} + e^{\frac{2\pi i}{3}}\ket{010} +
  e^{\frac{4\pi i}{3}}\ket{100}).
$$
Let $C_n$ denote the density matrix
$$C_n = \beta\sum_I C(I)C(I)^\dagger
$$ where $\beta$ is a normalizing factor so that $\tr(C_n)=1$. Observe
that $C_2$ is the density matrix $\ket{s}\bra{s}$ of the singlet state
$\ket{s}=\frac{1}{\sqrt{2}}(\ket{01}-\ket{10})$, so that the $C_n$ states
are $n$-qubit generalizations of the singlet.

Next we form products of $C_k$ states to make Werner states\footnote{It is perhaps nontrivial to show that the $C_n$ and the diagram states
constructed from them are indeed Werner states. This can be done with
straightforward calculations, but at the expense of technical
overhead. We refer the interested reader to our paper~\cite{symmmixed}
which gives details on the action of the Lie algebra of the local
unitary group on density matrices. One can show that the generators of
the Lie algebra of the Werner stabilizer group
$\Delta=\{(g,g,\ldots,g)\colon g\in SU(2)\}$ annihilate $C_n$.}. As with the
case of pure Werner states, we utilize diagrams. This time we consider
diagrams consisting of $n$ points labeled $1,2,\ldots,n$ on a
circle, with non-intersecting polygons that have vertices in the given set of
$n$ points. Again, there are a Catalan number $\frac{1}{n+1}{2n\choose n}$
of such $n$-vertex
diagrams\footnote{There is a
one-to-one correspondence between these ``non-intersecting polygon'' diagrams and the
``non-intersecting chord'' diagrams on $2n$ points in our pure
Werner states analysis of the previous section. Given a non-intersecting
chord diagram with $2n$ vertices, rename the vertices $1,1',2,2',
\ldots, n,n'$ and then glue each pair of $jj'$ for $1\leq j\leq n$.
}~\cite{stanleyenumcombvol2}.

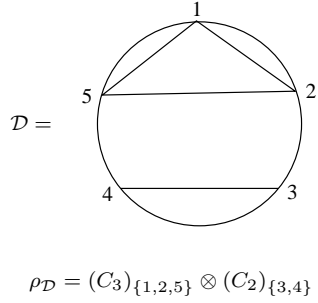
\begin{figure}
  \begin{center}
\input{diagramstate5.pstex_t}
\end{center}
\caption{\label{diagramstate5fig}A non-intersection polygon diagram state.}
\end{figure}

Given an $n$-vertex non-intersecting polygon diagram ${\cal D}$, we
construct a state $\rho_{\cal D}$ 
$$\rho_{\cal D}= \bigotimes_{U\in {\cal D}} C_U
$$
where the tensor has positions specified by elements of the partition
${\cal D}$ and $C_U$ denotes the state $C_{|U|}$ in qubit positions in
$U$. Figure~\ref{diagramstate5fig} shows an example.

Here is our main conjecture.

{\bf Conjecture.} The states $\rho_{\cal D}$ form a basis for the
space of Werner states (in the larger space of real linear combinations
of Pauli tensors).

Again, representation theory says that we have the right dimension: the
1-qubit density matrix representation space ${\cal V}$ decomposes into
irreducible $SU(2)$-submodules as
$${\cal V} = \{\mbox{span of } \sigma_0\} \oplus \{\mbox{span of }
\sigma_1,\sigma_2,\sigma_3\} = \R^1 \oplus \R^3$$ where the $\R^3$
summand is isomorphic to the adjoint representation. The
complexification ${\cal V}_\C$ is isomorphic to $\C^1 \oplus \C^3
\approx \C^2\otimes \C^2$, which is 2-qubit pure state space. In
general, the complexification $({\cal V}^{\otimes n})_\C$ of $n$-qubit
density matrix space ${\cal V}^{\otimes n}$ is isomorphic to $2n$-qubit
pure state space, as $SU(2)$ spaces. Thus the real dimension of the
trivial summand for $n$-qubit density matrices is equal to the complex
dimension of the trivial summand for $2n$-qubit pure states, which is
the Catalan number $\frac{1}{n+1}{2n \choose n}$.  This establishes that
we only need to show that the diagram states are independent in order to
prove the conjecture.

We conclude with a conjecture regarding a precise statement about the
stabilizer subgroup of the local unitary group for our constructed
Werner states. 
The full stabilizer of a
Werner state $\rho$, that is, the set 
$$\Stab_\rho =\{g\in G\colon \rho = g\rho g^\dagger\}
$$ of all local unitary transformations that fix $\rho$, could be larger
than the subgroup $\Delta$ of the unitary group. For example, a diagram
state $\rho_{\cal D}$ is stabilized by the subgroup
$$\Delta_{\cal D} :=\prod_{U\in {\cal D}} \Delta_U$$
where $\Delta_U$ denotes the subgroup that consists of elements
$(g,g,\ldots,g)$ in qubits in $U$, and all other coordinates are the identity.

In~\cite{su2blockstates}, we give a criterion on the diagrams that
appear in the expansion of a pure Werner state for when the stabilizer
subgroup of a pure Werner state is precisely the subgroup
$\Delta$ of the local unitary group, and not larger. The criterion is that for
any bipartition of the set of qubits, there must be a diagram (with
nonzero coefficient) in the expansion of the given state that has a
chord with one end in each of the sets of the partition. Our final
conjecture is a generalization of this idea to the general mixed Werner
case.

Consider a poset lattice of partitions of $\{1,2,\ldots,n\}$ (we
consider all partitions, with and without crossing polygons) where
${\cal D}\leq {\cal D'}$ if ${\cal D'}$ is a subdivision of ${\cal
  D}$. The $n$-gon is the least element at the bottom and the
all-singleton diagram is the greatest element at the top of this
lattice. The non-crossing polygon diagram lattice is a
sublattice. There is a corresponding lattice of subgroups of local
unitary group $G$, where $H$ is less than or equal to $K$ in the partial
order if $H$ is a subgroup of $K$. The subgroup $\Delta$ is the least
element at the bottom and $LG$ at the top. A diagram ${\cal D}$
corresponds to the subgroup $\Delta_{\cal D}$ defined above.
We conjecture that 
$$\Stab_{\sum a_{\cal D}\rho_{\cal D}} = \glb \{\Delta_{\cal D}\colon
a_{\cal D}\neq 0\}=\bigcap_{{\cal D}\colon a_{\cal D}\neq 0}
\Delta_{\cal D}$$ where 
``glb'' denotes greatest lower bound
in the lattice.  This would give a picture criterion for when a Werner
state has the Werner stabilizer (and not a larger one).

\section{Summary and Outlook}

We have surveyed known results on the structure and local unitary
equivalence classification of Werner states for the special cases of
pure states and symmetric states. We have presented a diagram
based construction for the general case of mixed Werner states that
generalizes the ``sums of products of singlets'' construction known for
pure states. Finally, we conjecture that the general construction will
prove to be a basis for the Werner states, and that this basis will lead
to local unitary classification and a precise analysis of stabilizer
subgroups.

\noindent {\em Acknowledgments.}  
This work has been supported by National Science Foundation grant
\#PHY-0903690.

\end{document}

%% file: spillustration4.pstex_t
\begin{picture}(0,0)%
\includegraphics{spillustration4.pstex}%
\end{picture}%
\setlength{\unitlength}{3947sp}%
\begingroup\makeatletter\ifx\SetFigFont\undefined%
\gdef\SetFigFont#1#2#3#4#5{%
  \reset@font\fontsize{#1}{#2pt}%
  \fontfamily{#3}\fontseries{#4}\fontshape{#5}%
  \selectfont}%
\fi\endgroup%
\begin{picture}(3226,1907)(1395,-2229)
\put(4183,-2183){\makebox(0,0)[b]{\smash{{\SetFigFont{8}{9.6}{\rmdefault}{\mddefault}{\updefault}{\color[rgb]{0,0,0}$-\ket{0101} - \ket{1010}$}%
}}}}
\put(2044,-1814){\makebox(0,0)[b]{\smash{{\SetFigFont{8}{9.6}{\rmdefault}{\mddefault}{\updefault}{\color[rgb]{0,0,0}${\cal P}=\{\{1,2\},\{3,4\}\}$}%
}}}}
\put(2044,-2025){\makebox(0,0)[b]{\smash{{\SetFigFont{8}{9.6}{\rmdefault}{\mddefault}{\updefault}{\color[rgb]{0,0,0}$s_{\cal P} = \ket{0101} + \ket{1010}$}%
}}}}
\put(3972,-1814){\makebox(0,0)[b]{\smash{{\SetFigFont{8}{9.6}{\rmdefault}{\mddefault}{\updefault}{\color[rgb]{0,0,0}${\cal Q}=\{\{1,4\},\{2,3\}\}$}%
}}}}
\put(3972,-2025){\makebox(0,0)[b]{\smash{{\SetFigFont{8}{9.6}{\rmdefault}{\mddefault}{\updefault}{\color[rgb]{0,0,0}$s_{\cal Q} = \ket{0011} + \ket{1100}$}%
}}}}
\put(2255,-2183){\makebox(0,0)[b]{\smash{{\SetFigFont{8}{9.6}{\rmdefault}{\mddefault}{\updefault}{\color[rgb]{0,0,0}$-\ket{0110} - \ket{1001}$}%
}}}}
\end{picture}%

%% file: diagramstate5.pstex_t
\begin{picture}(0,0)%
\includegraphics{diagramstate5.pstex}%
\end{picture}%
\setlength{\unitlength}{3947sp}%
\begingroup\makeatletter\ifx\SetFigFont\undefined%
\gdef\SetFigFont#1#2#3#4#5{%
  \reset@font\fontsize{#1}{#2pt}%
  \fontfamily{#3}\fontseries{#4}\fontshape{#5}%
  \selectfont}%
\fi\endgroup%
\begin{picture}(1818,1842)(985,-2052)
\put(1881,-2006){\makebox(0,0)[b]{\smash{{\SetFigFont{8}{9.6}{\rmdefault}{\mddefault}{\updefault}{\color[rgb]{0,0,0}$\rho_{\cal D} = (C_3)_{\{1,2,5\}} \otimes (C_2)_{\{3,4\}}$}%
}}}}
\put(1000,-1018){\makebox(0,0)[b]{\smash{{\SetFigFont{8}{9.6}{\rmdefault}{\mddefault}{\updefault}{\color[rgb]{0,0,0}${\cal D}=$}%
}}}}
\end{picture}%